\documentclass[runningheads]{llncs}
\usepackage{graphicx} 
 
\usepackage[table]{xcolor}
 
\usepackage{url}

\usepackage{authblk}
 
\usepackage[ruled,vlined]{algorithm2e}
\usepackage{algorithmic} 
\usepackage{amssymb,amsfonts} 
\renewcommand{\algorithmcfname}{ALGORITHM}
\SetAlFnt{\small}
\SetAlCapFnt{\small}
\SetAlCapNameFnt{\small}
\SetAlCapHSkip{0pt}

\usepackage{pgf}
\usepackage{tikz} 
\usepackage{multirow}
\usepackage[threshold=0]{csquotes}
 
\begin{document}
 
\title{Spatio-Temporal Analysis of Concurrent Networks} 

\author{Heinz Schmidt\inst{1}\orcidID{0000-0001-6278-4793} \and
Peter Herrmann\inst{2}\orcidID{0000-0003-3830-466X} \and
Maria Spichkova\inst{1}\orcidID{0000-0001-6882-1444} \and
James Harland\inst{1}\orcidID{0000-0001-8640-5137} \and
Ian Peake\inst{1}\orcidID{0000-0003-2136-5714} \and
Ergys Puka\inst{2}\orcidID{0000-0002-4257-3232}}
\authorrunning{H. Schmidt et al.} 
\institute{RMIT University, Melbourne, Australia
\email{\{heinz.schmidt, james.harland, ian.peake, maria.spichkova\}@rmit.edu.au} \and
Norwegian University of Science and Technology (NTNU), Trondheim, Norway\\
\email{\{peter.herrmann,ergys.puka\}@ntnu.no}}
\maketitle              

\newcommand{\CASTeL}{CASTeL}
\newcommand{\tictac}{Tic-Tac-Toe}
\newcommand{\CASTeLogic}{{Concurrent Alliances Spatio-Temporal Logic}}

\newcommand{\Until}{\mathcal{U}}
\newcommand{\Wuntil}{\mathcal{W}}
\newcommand{\stb}[3]{^{#1#2\##3}}
\newcommand{\meet}{\mathcal{MEET}}
\newcommand{\dab}[1]{\langle\langle #1\rangle\rangle}
\newcommand{\trl}{\vartriangleright\vartriangleleft}
\newcommand\natz{\mathbb{N}}
\newcommand\intz{\mathbb{Z}_{\geq 0}}
\newcommand\ratz{\mathbb{Q}_{\geq 0}}
\newcommand\realz{\mathbb{R}_{\geq 0}}
\newcommand\natpos{\mathbb{N}_{>0}}
\newcommand\intpos{\mathbb{Z}_{>0}}
\newcommand\ratpos{\mathbb{Q}_{>0}}
\newcommand\realpos{\mathbb{R}_{>0}}
\newcommand{\cev}[1]{\reflectbox{\ensuremath{\vec{\reflectbox{\ensuremath{#1}}}}}}

\newcommand{\multiset}[1]{\natz^{#1}}
\newcommand*\circled[2]{\tikz[baseline=(char.base)]{
            \node[shape=circle,draw=#2,fill=#2,text=white,inner sep=2pt] (char) {\Small #1};}}

\newcommand{\maria}[1]{{\textcolor{blue!80!black}{Maria: #1}}}
\newcounter{notecounter}
\setcounter{notecounter}{0}
\newcommand\authnote[3]{
   \stepcounter{notecounter}
   \marginpar{#1 \arabic{notecounter}$\Rsh$}  {\em\small $\Lsh$\arabic{notecounter}.#2 [#3]}
                      }  

\begin{abstract}
Many very large-scale systems are networks of cyber-physical systems in which humans and autonomous software agents cooperate.
To make the cooperation safe for the humans involved, the systems have to follow protocols with rigid real-time and real-space properties, but they also need to be capable of making competitive and collaborative decisions with varying rewards and penalties. Due to these tough requirements, the construction of system control software is often very difficult. This calls for applying a model-based engineering approach, which allows one to formally express the time and space properties and use them as guidance for the whole engineering process from requirement definition via system design to software development.
Moreover, it is beneficial, if one can verify with acceptable effort, that the time and space requirements are preserved throughout the development steps.
This paper focuses on modelling spatio-temporal properties and their model-checking and simulation using different analysis tools in combination with the methods and tool extensions proposed here. To this end, we provide an informal overview of \CASTeL, our \CASTeLogic. \CASTeL\ is stochastic and includes real-time concurrency and real-space distribution. \\
~\\
\emph{Preprint. Accepted to the 21th International Conference on Mobile and Ubiquitous Systems: Computing, Networking and Services (MobiQuitous 2024). Final version to be published by Springer (In Press).} 
\end{abstract}

\section{Introduction}

Very Large Scale Collaborative (VLSC) systems comprise defence,  
logistics, healthcare, and transport systems (including networked and autonomous vehicles) as well as 
widely distributed cloud computing and social networks, which may include millions of customers. 
Many of these VLSC systems are {\em cyber-physical systems} (CPS) \cite{BennaceurEtal2019:ModellingCyberphysical}, 
others are {\em cyber-social systems} (CSS), and 
some are both~\cite{ahmed2015:SoftwareDefinedNetworkinga}. 
A CPS operates, monitors and controls physical infrastructure such as
power and transport networks or manufacturing sites.
A CSS involves humans with mobile devices and many configurations and preferences. 
Cooperative intelligent transport systems are examples of uniting CPS and CSS, which include vehicle-to-vehicle and vehicle-to-infrastructure communication and wearable mobile devices that combine automated  
agents with human decision-making. 
An increasing portion of VLSC software is generated using rapid development tools and, more recently, generative Artificial Intelligence. 
Such systems often exhibit a significant number of program errors, which require automatic detection, reporting, correction and other responses in the large, that have been well studied in the literature, see, e.g.,~\cite{AmanullahEtal2023:MisbehaviourInTransportMetastudy}.

Our approach is based on abstract concurrent behaviour modelling with Petri Nets (PNs) and their use for model-checking stochastic logic specifications in Computation Tree Logic (CTL) variants~\cite{ClEm:81}. These logics include notions of abstract players (agents), with and without reward/penalty and other utility cost structures. They have a long history of model-checking over diverse PN model classes. 
PNs are a versatile form of state-transition systems that are suited to model system behaviour on very different levels of abstraction.
For instance, one can use them to specify limited synchronisation capabilities provided by integer semaphores, train rail segments, or car traffic crossings with limited capacity.
 
The relationship between different classes of PNs and various temporal logics is well known, which means that PNs can be considered as abstract, logical, operational, and executable specifications.
In particular, via model-checking one can formally carry out various analyses such as verifying structural properties of the nets, proving purely logical inference on the level of specified formulae, or simulating dynamic behaviour by executing the net either exhaustively or stochastically, in search of counter-examples that violate a logical formula.
 
This paper builds on our prior work for model-based and architecture-aware analysis of cyber-physical systems using a combination of {\em dependent finite state machine} interface descriptions, based on state-machine decomposable Petri Nets \cite{schmidt2012towards}. 
Further, this work is based on core aspects of readability that we introduced in~\cite{schmidt2019towards}.

\textbf{Contributions:} This paper proposes 
a model-driven approach called {\em \CASTeL}, which is short for {\em \CASTeLogic}.
\CASTeL\ makes the modelling of decisions, tactics and strategies in {\em competitive and cooperative systems} possible. 
Moreover, we show how one can use model-checking to verify, whether \CASTeL-formulae are realized by Coloured Stochastic Petri Nets (CSPNs).  
Besides the traditional strengths of modularity and temporal concurrency of CSPNs, the approach includes {\em spatial} reasoning about the {\em enforceability of properties}.

\textbf{Outline:} The rest of the paper is organised as follows. 
Section~\ref{sec:scenarios} introduces a motivating scenario for the proposed approach, while Sect.~\ref{sec:castelinformal} informally highlights features of our \CASTeL\ specifications and PN models.  
Section~\ref{Sec:RelatedWork} provides an overview of related approaches. 
Finally, Sect.~\ref{sec:conclusions} summarises the paper and suggests future work.

\section{Motivating scenario} 
\label{sec:scenarios}
 
In this section, we introduce our motivating scenario, which is about context-aware communication protocols for cellular dead spot mitigation.
Vehicles cooperating with each other and their infrastructure to realize Intelligent Transportation Systems (ITS), often suffer from the presence of dead spots, i.e., areas with no mobile network coverage.
Dead spots are quite common on large land masses such as Australia and Canada, where very few people live outside larger population centres. 
One example in Australia is a 300+km portion of the Silver Highway in New South Wales (NSW) between Mildura and Broken Hill. 

To alleviate this problem, we created special communication protocols which use opportunistic mobile ad hoc networks between nearby vehicles in a dead spot to alleviate transmission delays, see~\cite{MePH:19,PuHe:IJCPS19,PuHT:20}.
If the vehicles have reasonable estimates of their dead spot exit times, the ad hoc networks make the transfer of messages to those vehicles possible, that are anticipated to regain connectivity fastest.
The protocols were simulated using the versatile traffic simulator SUMO~\cite{SUMO:18}. 
The simulations have shown that using the context-aware protocols and the ad hoc networks may reduce the average waiting times in a dead spot by more than 40\%~\cite{PuHe:21}. 
While various vehicular network technologies can be used to realise  ad hoc networks between vehicles, we prefer the popular WiFi Direct protocol~\cite{WiFiDirect}, available on most state-of-the-art mobile phones.
It allows us to communicate over a distance of up to 200 meters, which is sufficient in most cases, see~\cite{PuHe:21}. In this paper, we  
{\em assume} that the WiFi Direct protocol is correctly implemented, and focus instead on the {\em emerging properties} across large numbers of vehicles participating in {\em message exchanges}.

While the SUMO-based simulations revealed interesting results~\cite{PuHe:21}, they lack formal proofs of important spatiotemporal properties. 
These formal verification tasks will be outlined below.
We ignore the possibility of messages getting lost in transit, such as a car not meeting the predicted exit time due to a driver changing the route or taking a break. We leave this to more refined ad hoc network protocols  
of our future work.
For simplicity, we also abstain from modelling our more advanced Context-Aware Message Flooding Protocol (CAMFLooP), which uses copies of the same message in multiple vehicles to guarantee an optimal reduction of the delivery time~\cite{PuHe:21,PuHT:20}.
Instead, we restrict ourselves to a prior version in which only one copy of a message is kept in the system at a time.
In ad hoc networks, this message is transmitted to the network peer predicted to leave the dead spot first~\cite{MePH:19,PuHe:IJCPS19}.

\begin{figure}[tb]
    \centering 
    \includegraphics[width=0.8\textwidth]{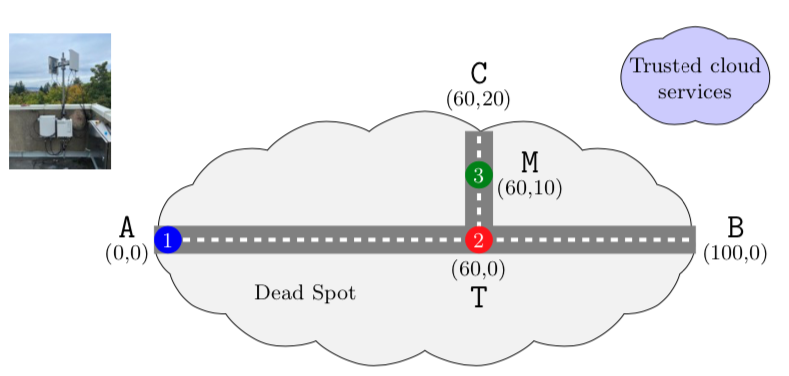}
    \caption{Simple Dead Spot Road Network with a T-intersection. 
    }
    \label{fig:RoadLayoutSimple}
\end{figure}

Figure~\ref{fig:RoadLayoutSimple} depicts a simple schematic dead spot, where $A$ and $B$ are two entry/exit points of the main road.
The points are at the coordinates $(0,0)$ and $(100,0)$, respectively. Further, the dead spot contains a T-intersection at point $T$ located at $(60,0)$, in which a side road from point $C$ at $(60,20)$ joins the main road. 
In this scenario, we assume the car's destination is known to the protocol in advance, like when we use Google or Apple Maps' ``Directions'' features. 
In the following, we consider through traffic on the routes $ATB$, $ATC$, $BTA$, $BTC$, $CTA$ and $CTB$.
The blue, red, and green circles in Fig.~\ref{fig:RoadLayoutSimple} outline the positions of three cars $c_1$, $c_2$, and $c_3$, that are all cut off from cellular network access in the depicted situation. 

\begin{figure}[tb]
    \centering
     \includegraphics[width=0.9\textwidth]{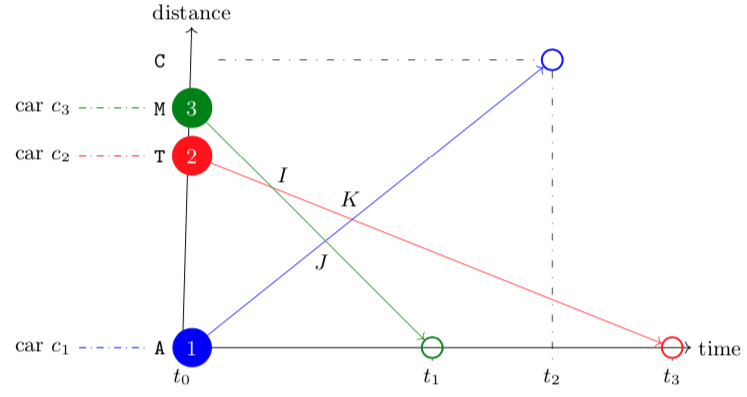}
    \caption{Simple scenario for the three cars shown in Fig. \ref{fig:RoadLayoutSimple}. 
    }
    \label{fig:Deadspot3cars}
\end{figure}

The trajectories of the three cars in our scenario are shown in Fig. \ref{fig:Deadspot3cars}. 
Here, car positions in the current situation (time $t_0=0$), i.e., the ones highlighted in Fig.~\ref{fig:RoadLayoutSimple}, are described by filled circles.
With hollow circles, we further designate the events in which vehicles leave the dead spot. We calculate the average speed through the dead spot from the time and relative distance, resulting in idealised linear car trajectories.  Car $c_1$ travels from $A$ to $C$ in time $t_2$. Car $c_2$ travels in the opposite direction from $T$ to $A$ in time $t_3$. Car $c_3$ entered at $C$ and travels to $A$ from its current point $M$ in the centre between $C$ and $T$, see Fig.~\ref{fig:RoadLayoutSimple}. 
The cars operate at different average speeds, represented by the gradients of the lines.
For instance, $c_3$ travels at a higher speed than $c_2$, which is overtaken at point $I$ between $T$ and $A$. At point $J$, $c_3$ is passing $c_1$, that runs in the opposite direction.

As described above, the dissemination protocol uses ad hoc networks to transmit messages between vehicles to speed their delivery~\cite{PuHe:IJCPS19}.
When two cars are in relative proximity (approximately at the intersection of their respective lines on Fig. \ref{fig:Deadspot3cars}), they may form an ad hoc network and exchange messages to improve the anticipated message delivery time. For example, a message jumping from the blue car $c_1$ to the green car $c_3$ at about $J$ (between $A$ and $T$) can be delivered at the earlier exit time $t_1$ of $c_3$ rather than the later time $t_2$ of $c_1$.
The message takes the effective route $AJA$.
Likewise, the ad-hoc networks created at points $I$ and $K$ can be used to shorten the time of message deliveries.

\section{Highlights of \CASTeL\ models and logical definitions} 
\label{sec:castelinformal}

\subsection{Petri Net model variations for the dead spot scenario}
\subsubsection{Discrete and continuous dynamic models}
The actual continuous trajectories of the cars $c_1$, $c_2$ and $c_3$ may deviate from the idealised linear trajectories depicted in Fig.~\ref{fig:Deadspot3cars}. We can imagine the accurate trajectories winding around the idealised average speed lines. If the changing speeds are known or stochastically generated in simulations, such a varying trajectory can be linearly approximated by a piecewise linear polygon. 
Each segment of this polygon represents the average speed of the given car on the corresponding road section. We envisage this by an equidistant segmentation of the $y$-axis of Fig. \ref{fig:Deadspot3cars}, i.e., the {\em spatial} division, rather than an equidistant division on the $x$-axis (time). This keeps with the interpretation of a Coloured Stochastic Petri Net (CSPN) as a compact discrete and continuous process generator. If the CSPN ignores the stochastic rates assigned to its transitions, one can analyse the abstract synchronization of its transitions by utilizing their net-induced partial order. The rates allow mapping any such partially ordered process onto a continuous timeline. If the specified rates are parameters in exponential distributions, the CSPN behaviour graph is equivalent to a continuous Markov chain or Markov decision process (if the CSPN is associated with reward structures), see~\cite{baarir2009greatspn}. Note that the behaviour graph of a Petri Net represents branching order or time. It forgets the spatial distribution and, therefore, the concurrency of transition firings.

\begin{figure}[tb]
    \centering
     \includegraphics[width=0.9\textwidth]{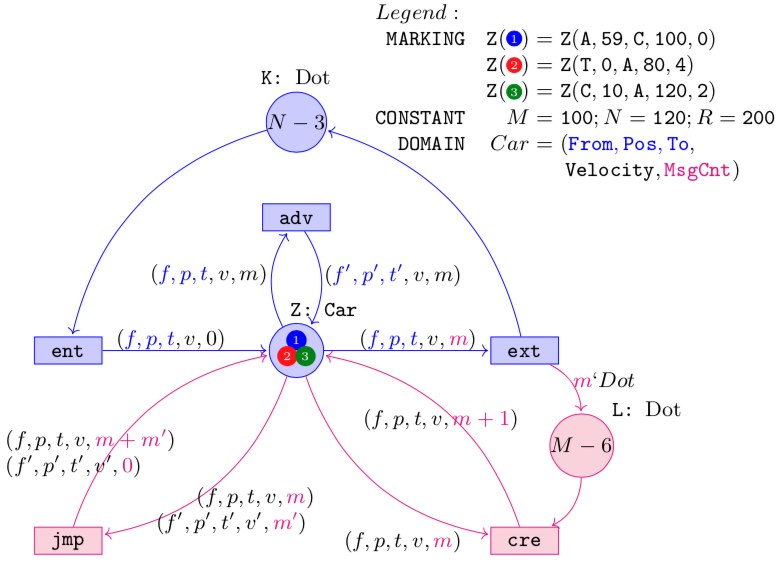}
    \caption{Coloured Stochastic Petri Net for concurrent car interactions in the simple dead spot network of Fig. \ref{fig:RoadLayoutSimple}. 
    }
    \label{fig:SPNsimplezone}
\end{figure}
Figure \ref{fig:SPNsimplezone} shows parts of a CSPN for the dead spot net in Fig.
\ref{fig:RoadLayoutSimple}.
To describe segments of the roads in our scenario, it uses the concept of {\em zones}.
For simplicity, we apply the distance of a road segment to the central point $T$, see Fig.~\ref{fig:RoadLayoutSimple}, to denominate a zone.
For example, car $c_1$ which is at the coordinate $(1,0)$, is $59$ units away from $T$ (coordinate $(60,0)$) and therefore in zone $59$.
The zones allow us to describe spatial properties, e.g., where on the road a car is or if two cars are nearby such that they can build up an ad hoc network.

Like in most PN descriptions, circles represent places and rectangles transitions.
The places $\tt K$ and $\tt L$ are marked by regular tokens (colour type {\em Dot}). 
$\tt K$ is used to bound the number of cars allowed in the dead spot, while $\tt L$ restricts the number of messages, the cars in the dead spot may carry at maximum.
In this way, it is guaranteed that our CSPN specifies only a finite number of different system states.
Place $\tt Z$ contains coloured tokens representing the cars currently in the dead spot.
Its tokens are marked by tuples from type $Car$, that have the form 
$(f,p,t,v,m)$ 
and include the following elements:
\begin{itemize}
    \item The entrance point $f\in\{A,B,C,T\}$, i.e. the point a car is coming from.
    \item 
    $p \in \mathbb{N}$ is an integer denoting the zone, the vehicle is currently in.
    \item 
    The exit point $t\in \{A,B,C\}$, i.e., the point it is heading to.
    \item 
    $v \in \mathbb{N}$ is the velocity of the car.
    \item $m \in \mathbb{N}$ lists the number of messages, the car carries.
\end{itemize}

In the marking shown in Fig.~\ref{fig:SPNsimplezone}, place $\tt L$ has $M-6$ tokens. $M$ is a constant model parameter describing the maximum number of messages possible in the system.
Thus, the tuple element $m$ of a token in place $\tt Z$ cannot exceed the value $M$, i.e., $0 \leq m \leq M$.
Another constant model parameter is $N$, that bounds the number of cars allowed in the dead spot, i.e., the number of tokens in place $\tt Z$. In the marking depicted in Fig.~\ref{fig:SPNsimplezone}, $\tt Z$ has three colour tokens of type $Car$.
In accordance with that, place $\tt K$ has $N-3$ tokens, and 
More precisely, the current marking $Mrk$ (during this hypothetical simulated run) includes $Mrk(L)=M-6$ and $Mrk(K)=N-3$, while $Mrk(Z)=\{(A,59,C,100,0),(T,0,A,80,4),(C,10,A,120,2)\}$ contains three colour tokens of type $Car$. This marking is reached after the three cars $c_1$ (blue), $c_2$ (red) and $c_3$ (green) entered the dead spot by executing transition $\tt ent$ and then advanced approximately to the situation depicted in Fig.~\ref{fig:Deadspot3cars} (transition $\tt adv$). Moreover, cars $c_2$ and $c_3$ created some message payload  (transition $\tt cre$). For each car entering the dead spot, the car capacity $\tt K$ shrinks by 1. Likewise, the message capacity $\tt L$ shrinks by 1 for each message created in the dead spot. As a car exits the dead spot (transition $\tt ext$), these capacities grow accordingly.

\begin{table}[tb]
    \centering
    \caption{Guards and rates for the transitions in Fig. \ref{fig:SPNsimplezone}.}
    \vspace{-3mm}
   \begin{tabular}{|l|l|l|}
   \hline  
   {\bf Transition} & {\bf Guard} & {\bf Rate} \\\hline\hline
   {\tt ent} & $IsRoute(f,t) \wedge f\not = T \wedge p = START(f)  $ & 1 \\\hline
   {\tt ext} &  $IsRoute(f,t) \wedge f = T \wedge p = START(t)$ & 1 \\\hline
   {\tt adv} &  
   $\begin{array}{rl}
     IsRoute(f,t) \wedge & \\
     ((f\not = T\wedge & \\
       ((p\not=0 &{\Rightarrow}\  p'=p-1\wedge f'=f) \: \vee \\
       (p=0 &{\Rightarrow}\  p'=0 \wedge f'=T)))\: \vee \ \\
     (f=T\wedge & p\not=START(t)\ \wedge \\
      f' = f & \wedge\ p'=p+1\wedge t'=t)) 
   \end{array}$ & 
   $0.04 \cdot v$  
   \\\hline
   {\tt cre} &  $true$    & 3 \\\hline
   {\tt jmp} & $ \begin{array}{l}IsClose(f,p,t,f',p',t') \ \wedge \\
   \ \ \ \ \ \ \ \ ETA(f,p,t,v) < ETA(f',p',t',v')
   \end{array}$ & 5
 \\\hline%
\end{tabular}
    \label{tab:guardratetable}
\end{table}

\subsubsection{Logical constraints and stochastic rates:}
Table \ref{tab:guardratetable} shows the guards and rates of the transitions used for the transitions in Fig.~\ref{fig:SPNsimplezone}.
In CSPNs, guards can be interpreted dynamically as transition firing conditions or statically as conditions for unfolding the coloured net into a basic uncoloured net. A fundamental transition exists only for colour value combinations that satisfy the transition guard in the basic net. In the coloured net, the higher-level transition fires only if the required input tokens are available and their values satisfy the guard, i.e., they generate an identical reachability graph on the same reachable markings.

Moreover, we use some additional predicates and functions: $IsRoute(f,t)$ is a predicate that holds if %
$f$ and $t$ are different points and $t$ is an exit point of the dead spot.
The function $START(f)$ indicates the initial zone position at a point $f$.
The predicate $IsClose(f,p,t,f',p',t')$ is $true$ if and only if the positions of the car with the tuple colours $f$, $p$, and $t$ and of the one characterised by $f'$, $p'$, and $t'$ are nearby. This accounts for cars driving in different directions, currently at different positions $p\not=p'$ but still in direct wifi range, and so on.
Finally, we use the function $ETA$ to express the expected time of arrival of a vehicle.

Rates of CSPNs are either probabilistic or they are real-valued parameters in probability density functions, for example, negative exponential functions $E(c,r,t)= c\cdot r\cdot e^{-r \cdot t}$ over time $t$ and rate $r$, with a scaling factor $c$. 
The rate function of the transition $\tt adv$ depends on the velocity $v$ of a car.

\subsubsection{Parameterized component-based architecture of models:}

In the initial marking (not shown in the figure), only place $\tt K$ is marked with $N$ tokens and place $\tt L$ with $M$ tokens. 
The current marking, depicted in the figure, is reachable from the above initial net marking.
The constants $M$ and $N$ bounding the numbers of messages and the vehicles in the dead spot, respectively, are definable model parameters.
Likewise, we use a constant $R$, that expresses the number of zones into which the roads in our scenario is segmented in.
Thus, our models are parameterised and give rise to a combinatorial family of nets over the possible range of these parameters. Parallel parameter sweeps \cite{YusufImanI.2015CRCa} may result in different simulation runs with a series of stochastic plots of various model performance measurements, such as statistical frequencies of combinations of states, transitions or formulae.

CSPNs are compositional. Linear algebra methods, including products, sums, scaling, and various other operations, have been proven to underpin their composition.
These operations on CSPNs can be visualized as merging the corresponding component nets by identifying some places or transitions. In Fig.~\ref{fig:SPNsimplezone}, we illustrate this by highlighting two net components:
\begin{itemize}
\item 

The blue subnet includes transitions for entry ($\tt ent$) to and exit ($\tt ext$) from the dead spot, respectively, or advancing ($\tt adv$) within a zone and between adjacent zones. 
    \item The magenta subnet is an extension adding messaging transitions, allowing the creation ($\tt cre$) of messages incrementing the message number of a car in transit; the jumping ($\tt jmp$) of messages between cars in proximity transfers all messages of one car to another vehicle if that car can reach its exit faster. 
\end{itemize}
As mentioned above, the properties of the dead spot mitigations protocols have already been modelled and simulated in SUMO \cite{KrajzewiczDaniel2010TSwS} in long runs, sometimes running on several workstations or virtual machines for weeks~\cite{PuHe:21}.
The protocols may be extended by further features like considering the reputation of drivers (agent) with respect to keeping their planned routes and speeds or their reliability in delivering the messages after having left the dead spot.
This reputation can then be used to decide which vehicle is the most reliable target for a given message. 
Furthermore, one can assume that some communication subscriptions support micro-payments, data entitlements, or other gratifications for drivers, so that trustworthy behaviour will also be rewarded financially. Thus, the modelling and simulation may include game-theoretic elements, where agents make ad hoc decisions regarding speed or destination with rewards in mind. Another variation provides modelling cyberattacks, where cars are hacked and behave maliciously, e.g., by offering their peers fake improvements in the message delivery when, in fact, they are discarding messages. Protocol variations offering increased fault tolerance in the presence of such attacks are of growing interest. 

This raises the question of how \CASTeL\ can feature component and product-line variation. In addition to parameters like $M$ and $N$, colour variations are an essential means of pa\-ram\-etrisation. For example, the combinatorial complexity of a CSPN may be managed by starting from simple models with a very limited enumeration or range type.
The model can then be step-wise refined by increasing the cardinality of the colour domain, i.e., by values that change the outcome of guard evaluations. Likewise, a colour product domain may be restricted using some formulae.
Conversely, conservative extensions correspond to injections. They permit some simulation  between a component model and the corresponding subnet or partial composition of the system as a whole. These colour restrictions and extensions correspond to projections and injections regarding nets and their runs. Parameter variations  may change a component's qualitative and quantitative interface or execution behaviour characteristics. These may result in the inclusion or exclusion of modules in the incremental or full integration of the system. 

\subsubsection{Gaps:} As mentioned above, CSPN modelling and simulation have been studied in the literature, and several education and industry-strengths tools exist.
However, our spatial interpretation in the \CASTeL\ context requires extensions, explicitly identifying spatial dimensions in these nets. For example, the highlighted blue colour variables $f$, $p$, and $t$ (see Fig. \ref{fig:SPNsimplezone}) represent abstract co-ordinates in terms of distances and zone points. Thus, they are {\em spatial variables} modelling the spatial aspects in the model of the dead spot mitigation protocol.

In Petri Nets, conflicts are branches in places where sufficient tokens enable multiple transitions, which can fire mutually exclusively. In coloured nets, conflicts are resolved in one of two ways. Either the information is already contained in the marking of the net, for example, the colour values of tokens. Or conflicts are non-deterministic since information enters the system from outside, for instance, by player choices, when we use the variant with alliances sketched above. In \CASTeL, player labels are particular colours, which may occur in tokens. According to the rules of the game, these are then assigned to the conflicts.
A conflict between transitions must only be resolved by the players assigned to it.

\subsection{Model design and logical requirement elicitation in \CASTeL}
The CSPN shown in Fig.~\ref{fig:SPNsimplezone} is a simplified discrete and abstract model for a variant of previously studied protocols~\cite{MePH:19,PuHe:IJCPS19}.  
We assume that each car $c_i$ enters with a random speed $v_i$ chosen from a few discrete values (e.g., 80--120 km/h).
It is easy to extend this variant to include variable speeds if desired. 
This can be easily realized by assuming, that the speed $v$ of a given car is constant in a zone, but that the transition {\tt adv} may change $v$ according to simulated choices of drivers or probabilistic events abstracting from traffic uncertainties. 
In consequence, the vehicle may have varying speeds in the different zones, it passes.

Based on this simple formal model, that includes the creation of messages by car passengers, by IoT components for machine-to-machine (M2M) communication, or logistic tracking,
we may wish to express its benefit by a precise specification, such as: 

\blockquote{\em At least $p$ percent of the messages created in the dead spot
arrive
in less than half the time, it would take the generating car to reach the planned exit of the dead spot. Suppose furthermore $1\leq n<10$ cars out of $10$  have satellite connection and offer forwarding of messages from other cars. This improves the average time to delivery by a factor $A\cdot n +B$, where $A$ and $B$ are constants determined by simulation runs.}

Our logic \CASTeL\ permits us to express claims like this one, including necessary spatial and temporal constraints, in an unambiguous and precise formalism. 
\CASTeL\ draws elements of its syntax from different CTL-based logics and extends and reconciles them through its use of extended Generalised Stochastic Petri Nets (GSPNs) as model generators for both state-transition models and their interpretations.
The use of GSPNs allow us to combine probabilistic chance with non-deterministic agent choice and reward. They are more general than models underlying, for example, ATL (Alternating-Time Temporal Logic, see \cite{alur2002alternating}) and rPATL (Probabilistic Alternating-time Temporal Logic with Rewards, see \cite{chen2013automatic}). In particular, stochastic rates may be real-valued time and space distributions (typically parameters in exponential distributions), for which Markovian probabilistic constraints may hold only under further constraints and normalisation. At the same time, non-Markovian PNs can be simulated statistically. PN models are not limited to alternating agent choice but can be truly concurrent.
This differs from interleaving models, where concurrency is specified by allowing two agents to operate in arbitrary order but not in parallel.
That is an important differentiator, especially in widely distributed systems and their performance analysis, simulation and estimation.

While the work here aims at model-checking and simulation modulo theory, our goal is also to lift necessary spatial abstractions onto an equal footing with real-time abstractions in logic based on PCTL. Such logics use bounded path formulae, in particular a logical {\em until} and {\em unless} operator with temporal bounds~\cite{HANSSONH1994Alfr}.
Such a bound limits the number of clock ticks for which a model checker checks whether safe conditions from states that might lead to hazards can be reached. In \CASTeL\, we therefore also permit spatial bounds, which count discrete spatial steps, i.e. routes that allow mobile actors to reach spatial goals, including a safe spatial region, and reap corresponding rewards. 
This can be realized by bounding firing sequences $t^n$ of a dangling stochastic transition $t$ stochastically by the rate of $t$, while in the underlying place-transition nets (ignoring the rates), they will be unbounded.
An equivalent completion (closure) is possible by adding a place $p_t$ initialised with $n$ tokens.
For example, we use the place {\tt K} representing the maximal number $M$ of cars in the dead spot for this purpose. 
In this way, we guarantee a hard bound
in the underlying net beside the soft bound by rates of the stochastic net.

Specifications like the one listed above make assumptions such as the following:
\blockquote{\em With a high probability (say 99\%), vehicles passing each other communicate reliably while in a range of 200 meters.}
Constraints are relevant for model realism, on the one hand, as we may need to exclude degenerate cases of dead spot models or car behaviour. On the other hand, constraints can avoid combinatorial explosions and unnecessary simulations of borderline cases, such as times of the day or week, when all cars travel in total isolation, times of major construction work or other hurdles, for which the protocol is not designed, or for which historical traffic data is not available. 
Principally, there are two kinds of constraints: 
\begin{enumerate}
    \item {\em Qualitative} constraints using different premises. For these, our improvement formulae become conclusions. Thus, the models and theories effectively become qualitative scenarios with different or varying surrounding conditions.
    \item {\em Quantitative} constraints. Here, we use stochastic model parameters. These vary stochastic rates in \CASTeL\  formulae. We may be able to plot behavioural observations or secondary stochastic observations as functions of varying parameter values; the variations may be discrete or continuous. We can also plug these varying parameters into the Petri Net simulation and look at the resulting variation of Petri Net properties and its provable or observed quantitative behaviour across various simulation runs.
\end{enumerate}

Let us make this a little more concrete by looking at the example of ``{\em coverage bubbles}'' --- groups of at least two collaborating cars temporarily forming an ad hoc network. In the terminology of game-theoretic logic like the Alternating Temporal Logic (ATL)~\cite{alur2002alternating}, such collections are called {\em alliances of players}, i.e., mutually independent decision makers. 
To deal with mobility, \CASTeL\ alliances are ad hoc. They form dynamically in space depending on the distance between the cars.
Moreover, the vehicles operate independently, i.e., their messaging process is automated, and they move concurrently. For example, multiple pairs of cars in one or more bubbles may exchange messages while moving in and out of proximity. Likewise, players in \CASTeL\ can make concurrent choices, such as changing their speed, adapting to traffic situations dynamically, etc. 
Regarding bubble formation, \CASTeL\ assumptions may include the following: 

\blockquote{\em With a significant minimal probability, say $20\%$, the spatial density of cars allows for the formation of bubbles of at least $3$ cars somewhere in the space of the dead spot.
}

On the one hand, considering such assumptions in CSPN simulations limits the scenarios to be analysed by simulation runs, exploring possible processes and their stochastic properties.
On the other hand, the assumptions simplify model-checking of logical guarantees required as part of requirements definitions.

\begin{figure}[tb]
    \centering
     \includegraphics[width=0.9\textwidth]{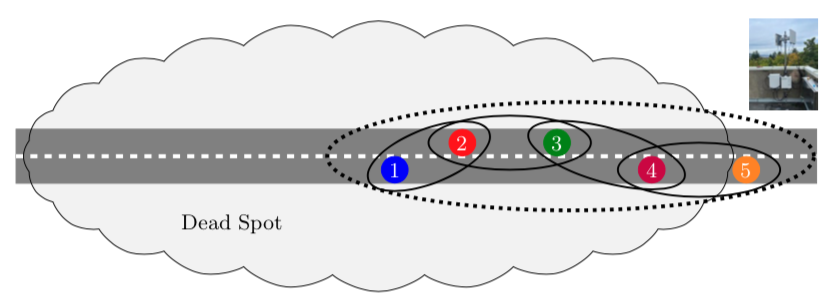}
    \caption{Scenario with a coverage bubble formed from ad hoc networks between vehicles. 
    }
    \label{fig:ConnectedGraph}
\end{figure}

Stochastic rates of car arrivals at entry points can set time intervals at the same space point, i.e., bounds for the average space-time density across a dead spot.
Following the concept of Multi-hop Cellular Network (MCN)~\cite{LiHs:00}, in which a mobile unit communicates with a fixed base station utilizing others as relay stations, the proximity relation between cars may define a connected graph for a crowded dead spot.
This graph is the {\em support} of the coverage bubble. Messages inside a coverage bubble can exit by a sequence of hops at a speed limited by communication speed and the number of hops required.
An example scenario is depicted in Fig.~\ref{fig:ConnectedGraph}.
Here, the ad hoc network connections between the five vehicles $c_1$ to $c_5$ (solid ellipses) form a coverage bubble, which is visualized as a dotted ellipse.
This makes it possible that, e.g., messages stored in $c_1$ can be forwarded via some other vehicles acting as relay stations to $c_5$, which is outside the dead spot and can forward the messages via the cellular network.

The time of a single hop is dominated by the time it takes to establish a connection or switch from one to another. Furthermore, it depends on message lengths. Since the density of cars may vary, coverage bubbles may not connect directly to the dead spot exit.
Moreover, a bubble may disintegrate fast when its support cars travel in opposing directions, e.g., in Fig.~\ref{fig:ConnectedGraph} cars $c_2$ and $c_3$  move in the opposite direction of all the other cars and will soon be out of range.
Then, $c_1$ cannot form a bubble with $c_4$ and $c_5$ anymore. 
However, a group of vehicles travelling in the same direction at similar speeds may maintain a coverage bubble for an extended period,
e.g., $c_2$ and $c_3$.

\subsubsection{Gaps}
Like for models, the \CASTeL\ logical specification requires extensions to the logics it builds upon.
Firstly, our PCTL-like formulae use a bounded {\em until} operator. While this is sufficient for iterative time bounds, \CASTeL\ also needs spatial bounds for iterating over space.
For example, when reasoning inductively about vehicle routes, one may specify that cars will reach their planned destination based entirely on the road network's structure and spatial decisions.
Secondly, players in alliances may resolve conflicts concurrently, and reward structures may be associated with alliances or individual players.
Thirdly, alliances may be ad hoc. For example, bubbles of vehicles in proximity form such alliances. To this end, ATL-like formulae using the enforceability operator must be allowed to range over ad hoc alliances. To this end, we use player sets constrained by predicates like the $IsClose$ predicate used in Table \ref{tab:guardratetable}.

\section{Related Work}
\label{Sec:RelatedWork}

Stochastic reward PNs have been used prior in the work of Chiola et al.~\cite{Chiola85} and Marsan et al.~\cite{ajmone1985petri}. 
Their main objective was modelling close to the domain-specific informal models of practitioners, their quantitative reliability estimates, and 
risk management requirements. However, while methods to specify external events and risks were included, those to describe security risks such as cyber-attacks were not.
The approach formed a basis for the creation of the GreatSPN tool for the stochastic analysis of systems modelled as (stochastic) PNs, see
\cite{amparore201630,chiola1995greatspn}. 
Heiner et al. introduced a comparative study of stochastic analysis techniques in \cite{heiner2010comparative}. 
Part of their approach was the tool \emph{Snoopy} to model and simulate hierarchical
graph-based system descriptions, see~\cite{heiner2008snoopy}. It allows users to analyse several kinds of PNs, including timed PNs and stochastic PNs.

The probabilistic model checker \emph{Storm} was introduced in \cite{dehnert2017storm,hensel2022probabilistic}. Storm supports the analysis of discrete- and continuous-time variants of both Markov chains and Markov decision processes (MDPs). 
Open-source Storm methods and tools are used in the backend of GreatSPN and PRISM.

Spatio-temporal models for formal analysis and property-based testing were presented in \cite{alzahrani2016spatio,spichkova2014modeling}. These works focus on supporting software engineers in understanding temporal models (making formal representations more accessible to the industrial application). Moreover, they provide a schematic translation of time-based constructs to the spatial analyser BeSpaceD~\cite{BlSc:14}. 
Another approach to model spatial aspects of safety-critical systems was presented in \cite{spichkova2014modeling}. That work applies the formal language Focus\textsuperscript{ST} and presents an example system based on interacting autonomous vehicles.

In our previous work on BSpaceD  \cite{herrmann2017:FormalAnalysisControl,herrmann2019model}, we have used theorem-proving modulo theory based on the Microsoft SMT prover Z3 \cite{Bjoerner2023:Z3SMT} applied to industrial robotics and autonomous vehicles, such as collision avoidance in a primarily automated fulfilment centre. Space-time-related algorithms were written in Scala and check the satisfiability of speed-dependent safe spatiotemporal separation of any pair of mobile autonomous objects in proximity, including robot-to-robot movement or mobile robots in safe separation from human staff and movement.
%

\section{Conclusion and Future work}
\label{sec:conclusions} 

This paper presented an overview of \CASTeL, a \CASTeLogic. We illustrated it by a motivating scenario, in which we model an opportunistic mobile ad-hoc network formed by the cars or trucks crossing a communication dead spot and analyse possible variations.

The \CASTeL\ approach builds on the marriage of elements from PCTL and ATL with reward structures and model-checked using concurrent Petri Nets, particularly Coloured Generalised Stochastic Petri Nets (GSPN). \CASTeL\ stands apart from other temporal logics by providing extensions for spatial abstractions and spatially stochastic distributions. The concurrent interpretation of players (decision makers) enables reasoning and analyses about the collaboration within alliances and competition between them. Furthermore, the modelling focuses on system-level properties, which emerge from the interaction of the system components in the context of many concurrent individual failures, misbehaviours or malicious actions.
We motivated these novel extensions with a mobile ad hoc network example and showed how dynamic alliance formation can be modelled.
Moreover, we outlined how emergent safety and security properties can be guaranteed by model-checking within stochastic bounds using existing tools or suggested extensions. 
Our work uses backend tools with well-documented scalability limitations, in particular stochastic model-checkers and Petri net tools.

\textbf{Future Work:} 
Our future work will focus on  detailed presentation of  the \CASTeL\ semantics and performance evaluations to help prioritise backend extensions aiming to relax limitations.
One of the directions to evolve \CASTeL\ is to cover scenarios where speeds vary arbitrarily by player choices, potentially due to road or traffic conditions. 
We already modelled the spatio-temporal resolution using model parameters (constants) and can therefore generate a sequence of models of finer granularity and plot the increasingly refined approximations of continuous models.
Another direction is to consider the behaviour of drivers in an ad hoc alliance, and to classify them into categories such as cooperative, uncommitted or adversarial, corresponding to their willingness to deliver messages when exiting dead spots. This may include consideration of the relative rates of each type, such as in say 20 vehicles, there are expected to be 12 cooperative drivers, 7 uncommitted and 1 adversarial. In addition, one could consider the relative probabilities of cooperative, uncommitted and adversarial behaviour across sparse versus dense areas, or according to the number of vehicles in a given area.  

In a future refined model, (i) the average speed may vary during a simulation run, for example, by the hour of the day and the day of the year, and (ii) the speed at a location may be varied for different cars, by adding additional token colours to a car (such as some proxy for speed: slow, moderate, fast or speeding). 

Recent research has used machine learning to determine or fine-tune the rates in stochastic Petri Nets. About \CASTeL\ and its underlying CSPN models, there are two promising approaches. 
Firstly, machine learning may assist in reducing the complexity of the reachability problem \cite{QiHongda2023PRPo}, permitting relatively fast simulation and model-checking by trading off accuracy about rare events.
Secondly, AI learning of Markov processes has been applied to stochastic Petri Nets based on their reachability graphs when these are Markovian~\cite{MaoHaoyang2023Adsp,VansonGautier2022EDmi}. Then, machine-learning methods for Markov models and decision processes can be applied directly. However, non-Markovian reachability graphs reflect some history sensitivity. Large language models may capture sufficient history in the current state, i.e. via additional states and weights of their outgoing transitions.

\bibliographystyle{splncs04}

\end{document}